\documentclass[review]{elsarticle}

\usepackage{multirow}
\usepackage{hyperref}
\usepackage{float}
\usepackage{ccaption}

\begin{document}

\begin{frontmatter}

\title{Exploring Chemical Space using Natural Language Processing Methodologies for Drug Discovery}

\author[cmpe]{Hakime \"{O}zt\"{u}rk}
\author[cmpe]{Arzucan \"{O}zg\"{u}r}
\author[IBM]{Philippe Schwaller}
\author[IBM]{Teodoro Laino \corref{mycorrespondingauthor}}
\ead{teo@zurich.ibm.com}
\author[che,zurich]{Elif Ozkirimli \corref{mycorrespondingauthor}}
\ead{elif.ozkirimli@boun.edu.tr}
\ead{+41 76 349 7471}
\cortext[mycorrespondingauthor]{Corresponding author}

\address[cmpe]{Department of Computer Engineering, Bogazici University, Istanbul, Turkey}
\address[IBM] {IBM Research, Zurich, Switzerland}
\address[che]{Department of Chemical Engineering, Bogazici University, Istanbul, Turkey}
\address[zurich]{Department of Biochemistry, University of Zurich, Winterthurerstrasse 190, CH-8057 Zurich, Switzerland}


\begin{abstract}
Text based representations of chemicals and proteins can be thought of as unstructured languages codified by humans to describe domain specific knowledge. Advances in natural language processing (NLP) methodologies in the processing of spoken languages accelerated the application of NLP to elucidate hidden knowledge in textual representations of these biochemical entities and then use it to construct models to predict molecular properties or to design novel molecules. This review outlines the impact made by these advances on drug discovery and aims to further the dialogue between medicinal chemists and computer scientists.


\paragraph{Teaser} The application of natural language processing methodologies to analyze text based representations of molecular structures opens new doors in deciphering the information rich domain of biochemistry toward the discovery and design of novel drugs. 

\end{abstract}

\end{frontmatter}


\section{Introduction}

The design and discovery of novel drugs for protein targets is powered by an understanding of the underlying principles of protein-compound interaction. Biochemical methods that measure affinity and biophysical methods that describe the interaction in atomistic level detail have provided valuable information toward a mechanistic explanation for bimolecular recognition \cite{schneider2018automating}. However, more often than not, compounds with drug potential are discovered serendipitously or by phenotypic drug discovery \cite{moffat2017opportunities} since this highly specific interaction is still difficult to predict \cite{duarte2019integration}. Protein structure based computational strategies such as docking \cite{sledz2018protein},  ultra-large library docking for discovering new chemotypes \cite{lyu2019ultra}, and molecular dynamics simulations \cite{sledz2018protein} or ligand based strategies such as quantitative structure-activity relationship (QSAR) \cite{schneider2016novo, bosc2019large}, and molecular similarity \cite{eckert2007molecular} have been powerful at narrowing down the list of compounds to be tested experimentally. With the increase in available data, machine learning and deep learning architectures are also starting to play a significant role in cheminformatics and drug discovery \cite{lo2018machine}. These approaches often require extensive computational resources or they are limited by the availability of 3D information. On the other hand, text based representations of biochemical entities are more readily available as evidenced by the 19,588 biomolecular complexes (3D structures) in PDB-Bind \cite{wang2005pdbbind} (accessed on Nov 13, 2019) compared with 561,356 (manually annotated and reviewed) protein sequences in Uniprot \cite{apweiler2004uniprot} (accessed on Nov 13, 2019) or 97 million compounds in Pubchem \cite{bolton2008pubchem} (accessed on Nov 13, 2019). The advances in natural language processing (NLP) methodologies make processing of text based representations of biomolecules an area of intense research interest.

The discipline of natural language processing (NLP) comprises a variety of methods that explore a large amount of textual data in order to bring unstructured, latent (or hidden) knowledge to the fore \cite{manning1999foundations}. Advances in this field are beneficial for tasks that use language (textual data) to build insight. The languages in the domains of bioinformatics and cheminformatics can be investigated under three categories: (i) natural language (mostly English) that is used in documents such as scientific publications, patents, and web pages,  (ii) domain specific language, codified by a systematic set of rules extracted from empirical data and describing the human understanding of that domain (e.g. proteins, chemicals, etc), and (iii) structured forms such as tables, ontologies, knowledge graphs or databases \cite{oliveira2019leveraging}. Processing and extracting information from textual data written in natural languages is one of the major application areas of NLP methodologies in the biomedical domain (also known as \textit{BioNLP}). Information extracted with BioNLP methods is most often shared in structured databases or knowledge graphs \cite{ernst2015knowlife}. We refer the reader to the comprehensive review on \textit{BioNLP} by \citet{krallinger2017information}. Here, we will be focusing on the application of NLP to domain specific, unstructured biochemical textual representations toward exploration of chemical space in drug discovery efforts. 

We can view the textual representation of biomedical/biochemical entities as a domain-specific language. For instance, a genome sequence is an extensive script of four characters (A, T, G, C) constituting a genomic language. In proteins, the composition of 20 different natural amino acids in varying lengths builds the protein sequences. Post-translational modifications expand this 20 letter alphabet and confer different properties to proteins \cite{karve2011small}. For chemicals there are several text based alternatives such as chemical formula, IUPAC International Chemical Identifier (InChI)  \cite{heller2013inchi} and Simplified Molecular Input Line Entry Specification  (SMILES) \cite{weininger1988smiles}. 

Today, the era of ``big data" boosts the ``learning" aspect of computational approaches substantially, with the ever-growing amounts of information provided by publicly available databases such as PubChem \cite{bolton2008pubchem}, ChEMBL \cite{gaulton2011chembl}, UniProt \cite{apweiler2004uniprot}. These databases are rich in biochemical domain knowledge that is in textual form, thus building an efficient environment in which NLP-based techniques can thrive. Furthermore, advances in computational power allow the design of more complex methodologies, which in turn drive the fields of machine learning (ML) and NLP. However, biological and chemical interpretability and explainability  remain  among the major challenges of AI-based approaches. Data management in terms of access, interoperability and reusability are also critical for the development of NLP models that can be shared across disciplines.

With this review, we aim to provide an outline of how the field of NLP has influenced the studies in bioinformatics and cheminformatics and the impact it has had over the last decade. Not only are NLP methodologies facilitating processing and exploitation of biochemical text, they also promise an ``understanding" of biochemical language to elucidate the underlying principles of bimolecular recognition. NLP technologies are enhancing the biological and chemical knowledge with the final goal of accelerating drug discovery for improving human health. We highlight the significance of an interdisciplinary approach that integrates computer science and natural sciences.

\subsection{NLP Basics}
\citet{chowdhury2003natural} describes NLP on three levels: (i) the word level in which the smallest meaningful unit is extracted to define the morphological structure, (ii) the sentence  level where grammar and syntactic validity are determined, and (iii) the domain or context level in which the sentences have global meaning. Similarly, our review is organized in three parts in which bio-chemical data is investigated at: (i) word level, (ii) sentence (text) level, and (iii) understanding text and generating meaningful sequences. Table \ref{tab:nlpconcepts} summarizes important NLP concepts related to the processing of biochemical data. We refer to these concepts and explain their applications in the following sections. 


All NLP technology relates to specific AI architectures. In Table \ref{tab:methodologies} W-we summarize the main ML and deep learning (DL) architectures that will be mentioned throughout the review. 


\section{Biochemical Language Processing} \label{section:NLP4BioChem}

The language-like properties of text-based representations of chemicals were recognized more than 50 years ago by Garfield \cite{garfield1961chemico}. He proposed a ``chemico-linguistic" approach to representing chemical nomenclature with the aim of instructing the computer to draw chemical diagrams. Protein sequence has been an important source of information about protein structure and function since Anfinsen's experiment \cite{anfinsen1973principles}. Alignment algorithms, such as Needleman-Wunsh  \cite{needleman1970general} and Smith-Waterman \cite{smith1981identification}, rely on sequence information to identify functionally or structurally critical elements of proteins (or genes). 

To make predictions about the structure and function of compounds or proteins, the understanding of these sequences is critical for bioinformatics tasks with the final goal of accelerating drug discovery. Much like a linguist who uses the tools of language to bring out hidden knowledge, biochemical sequences can be processed to propose novel solutions, such as predicting interactions between chemicals and proteins or generating new compounds based on the level of understanding. In this section, we will review the applications of some of the NLP-concepts to biochemical data in order to solve bio/cheminformatics problems.

\subsection{Textual Chemical Data} \label{section:availabledata}

Information about chemicals can be found in repositories such as PubChem \cite{bolton2008pubchem}, which includes information on around 100 million compounds, or Drugbank \cite{wishart2006drugbank}, which includes information on around 10,000 drugs. The main textual sources used in drug discovery are textual representations of chemicals and proteins. Table \ref{tab:resources} lists some sources that store different types of biochemical information.  


Chemical structures can be represented in different forms that can be one-dimensional (1D), 2D, and 3D. Table \ref{tab:ampicillin} depicts  different identifiers/representations of the drug \textit{ampicillin}. While the 2D and 3D representations are also used in ML based approaches \cite{lo2018machine}, here we focus on the 1D form, which is the representation commonly used in NLP.

\paragraph{IUPAC name} The International Union of Pure and Applied Chemistry (IUPAC) scheme (i.e. nomenclature) is used to name compounds following pre-defined rules such that the names of the compounds are unique and consistent with each other (\url{iupac.org/}).  

\paragraph{Chemical Formula}
The chemical formula is one of the simplest and most widely-known ways of describing chemicals using letters (i.e. element symbols), numbers, parentheses, and (-/+) signs. This representation gives information about which elements and how many of them are present in the compound. 

\paragraph{SMILES}
The Simplified Molecular Input Entry Specification (SMILES) is a text-based form of describing molecular structures and reactions \cite{weininger1988smiles}. SMILES strings can be obtained by traversing the  2D graph representation of the compound and therefore SMILES provides more complex information than the chemical formula. Moreover, due to its textual form, SMILES takes 50\% to 70\% less space than other representation methods such as an identical connection table (\url{daylight.com/dayhtml/doc/theory/theory.smiles.html}). 

SMILES notation is similar to a language with its own set of rules. Just like it is possible to express the same concept with different words in natural languages, the SMILES notation allows molecules to be represented with more than one unique SMILES. Although this may sound like a significant ambiguity, the possibility of using different SMILES to represent the same molecule was successfully adopted as a data augmentation strategy by various groups (\citet{bjerrum2017smiles, kimber2018synergy, schwaller2018molecular}).

Canonical SMILES can provide a unique SMILES representation. However, different databases such as PubChem and ChEMBL might use different canonicalization algorithms to generate different unique SMILES. OpenSMILES (\url{opensmiles.org/opensmiles.html}) is a new platform that aims to universalize the SMILES notation. In isomeric SMILES, isotopism and stereochemistry information of a molecule is encoded using a variety of symbols (``/", ``\textbackslash", ``@", ``@@").

\paragraph{DeepSMILES} DeepSMILES is a novel SMILES-like notation that was proposed to address two challenges of the SMILES syntax: (i) unbalanced parentheses and (ii) ring closure pairs \cite{OBoyle2018}. It  was initially designed to enhance machine/deep-learning based approaches that utilize SMILES data as input (\url{github.com/nextmovesoftware/deepsmiles}). DeepSMILES was adopted in a drug-target binding affinity prediction task in which the findings highlighted the efficacy of DeepSMILES over SMILES in terms of identifying undetectable patterns \cite{ozturk2018chemical}. DeepSMILES was also utilized in a molecule generation task in which it was compared to canonical and randomized SMILES text \cite{arus2019randomized}. Here, the results suggested that DeepSMILES might limit the learning ability of the SMILES-based molecule generation models because its syntax is more grammar sensitive with the ring closure alteration and the use of a single symbol for branching (i.e. ``)") introducing longer sequences.

\paragraph{SELFIES} SELF-referencIng Embedding Strings (SELFIES) is an alternative sequence-based representation that is built upon ``semantically constrained graphs" \cite{krenn2019selfies}. Each symbol in a SELFIES sequence indicates a recursive Chomsky-2 type grammar, and can thus be used to convert the sequence representation to a unique graph. SELFIES utilize SMILES syntax to extract words that will correspond to  semantically valid graphs (\url{github.com/aspuru-guzik-group/selfies}). \citet{krenn2019selfies} compared SELFIES, DeepSMILES and SMILES representations in terms of validity in cases where random character mutations are introduced. The evaluations on the QM9 dataset yielded results in the favor of SELFIES. 

\paragraph{InChI} InChI is the IUPAC International Chemical Identifier, which is a non-proprietary and open-source structural representation (\url{inchi-trust.org}) \cite{heller2015inchi}. The InChIKey is a character-based representation that is generated by hashing the InChI strings in order to shorten them.  InChi representation has several layers (each) separated by the ``/" symbol.  

The software that generates InChi is publicly available and InChi does not suffer from ambiguity problems. However, its less complex structure makes the SMILES representation easier to use as shown in a  molecular generation study \cite{gomez2018automatic} and in building meaningful chemical representations with a translation-based system   \cite{winter2019learning}. Interestingly, the translation model was able to translate from InChi to canonical SMILES, whereas it failed to translate from canonical SMILES to InChi. \citet{winter2019learning} suggested that the complex syntax of InChi made it difficult for the model to generate a correct sequence. 

\paragraph{SMARTS} SMiles ARbitrary Target Specification (SMARTS) is a language that contains specialized symbols and logic operators that enable  substructure (pattern) search on SMILES strings \cite{ghersi2014molblocks}.  SMARTS can be used in any task that requires pattern matching on a SMILES string such as, querying databases or creating rule dictionaries such as RECAP \cite{lewell1998recap} and BRICS \cite{degen2008art} to extract fragments from SMILES (\url{daylight.com/dayhtml/doc/theory/theory.smarts.html}).

\paragraph{SMIRKS} SMIRKS notation can be used to describe generic reactions (also known as transforms) that comprise one or more changes in atoms and bonds (\url{https://daylight.com/daycgi\_tutorials/smirks\_examples.html}). These transforms are based on ``reactant to product" notation, and thus make use of SMILES and SMARTS languages. SMIRKS is utilized in tasks such as constructing an online transform database \cite{avramova2018retrotransformdb} and predicting metabolic transformations \cite{arvidsson2017prediction}. A recent study achieves a similar performance to rule-based systems in classifying chemical reactions by learning directly from SMILES text with transforms via neural networks  \cite{schwaller2019data}.

\subsection{Identification of Words/Tokens} \label{subsection:words}

Similar to  words in natural languages, we can assume that the ``words" of  biochemical sequences convey significant information (e.g. folding, function etc) about the entities. In this regard, each compound/protein is analogous to a sentence, and each compound/protein unit is analogous to a word. Therefore, if we can decipher the grammar of biochemical languages, it would be easier to model bio/cheminformatics problems. However,  protein and chemical words are not explicitly known and different approaches are needed to extract syntactically and semantically meaningful biochemical word units from these textual information sources (i.e. sequences). Here, we review some of the most common tokenization approaches used to determine the words of biochemical languages.

\paragraph{$k$-mers ($n$-grams)} One of the simplest approaches in NLP to extract a small language unit is to use $k$-mers, also known as $n$-grams. $k$-mers indicate $k$ consecutive overlapping characters that are extracted from the sequence with a sliding window approach. ``LINGO", which is one of the earliest applications of $k$-mers in cheminformatics, is the name of the overlapping $4$-mers that are extracted from SMILES strings \cite{vidal2005}. $4$-mers of the SMILES of \textit{ampicillin}, ``CC1(C(N2C(S1)C(C2=O)NC(=O)C(C3=CC=CC=C3)N)C(=O)O)C", can be listed as \{ `CC1(', `C1(C', `1(C(', ..., `O)O)', `)O)C' \}. From a sequence of length $l$, a total of $(l-n)+1$ $k$-mers can be extracted. Extracting LINGOs from SMILES is a simple yet powerful idea that has been successfully used to compute molecular similarities, to differentiate between bioisosteric and random molecular pairs \cite{vidal2005} and in a drug-target interaction prediction task \cite{ozturk2016comparative}, without requiring 2D or 3D information. The results suggested that a SMILES-based approach to compute the similarity of chemicals is not only as good as a 2D-based similarity measurement, but also faster \cite{ozturk2016comparative}. 

$k$-mers were successfully utilized as protein \cite{asgari2015continuous} and chemical words \cite{ozturk2018novel} in protein family classification tasks. $3$-mers to $5$-mers were often considered as the words of the protein sequence. \citet{motomura2012word} reported that  some $5$-mers could be matched to motifs and  protein words are most likely a mixture of different $k$-mers. For the protein function prediction task, \citet{cao2017prolango} decided to choose among the 1000 most frequent words to build the protein vocabulary, whereas \citet{ranjan2019deep} utilized each $k$-mer type separately and showed that $4$-mers provided the best performance. In the latter work, instead of using the whole protein sequence, the words were extracted from different length protein segments, which are also long $k$-mers (i.e. 100-mer, 120-mer) with 30 amino-acid gaps. The use of segmented protein sequences yielded better results than using the whole protein sequence, and important and conserved subsequences were highlighted. $k$-mers were also used as features, along with position specific score matrix features, in the protein fold prediction problem \cite{wei2015enhanced}.

\paragraph{Longest Common Subsequences} The identification of the longest common subsequence (LCS) of two sequences is critical for detecting their similarity. When there are multiple sequences, LCSs can point to informative patterns. LCSs extracted from SMILES sequences performed similarly well to $4$-mers in chemical similarity calculation  \cite{ozturk2016comparative}.

\paragraph{Maximum Common Substructure} \citet{cadeddu2014organic} investigated  organic chemistry as a language in an interesting study that extracts maximum common substructures (MCS) from the 2D structures of pairs of compounds to build a vocabulary of the molecule corpus. Contrary to the common idea of functional groups (e.g. methyl, ethyl etc.) being ``words" of the chemical language, the authors argued that  MCSs (i.e. fragments) can be described as the words of the chemical language \cite{cadeddu2014organic}.  A recent work investigated the distribution of these words in different molecule subsets \cite{wozniak2018linguistic}.  The ``words" followed \textit{Zipf's Law}, which indicates the relationship between the frequency of a word and its rank (based on the frequency) \cite{zipf1949human}, similar to most natural languages. Their results also showed that drug ``words" are shorter compared to natural product ``words".

\paragraph{Minimum Description Length} 
Minimum Description Length (MDL) is an unsupervised compression-based word segmentation technique in which words of an unknown language are detected by compressing the text corpus. In a protein classification task, each protein was assigned to the family in which its sequence is compressed the most, according to the MDL-based representation \cite{ganesan2017protein}.  \citet{ganesan2017protein} investigated whether the MDL-based words of the proteins show similarities to PROSITE patterns \cite{hulo2006prosite} and showed that less conserved residues  were compressed less by the algorithm.  \citet{ganesan2017protein} also emphasized that the integration of domain knowledge, such as the consideration of the hydrophilic and hydrophobic aminoacids in the words  (i.e. grammar building), might prove effective. 

\paragraph{Byte-Pair Encoding} 

Byte-Pair Encoding (BPE) generates words based on high frequency subsequences starting from frequent characters \cite{sennrich2015neural}. A recent study  adopted a linguistic-inspired approach to predict protein-protein interactions (PPIs) \cite{wang2019high}.  Their model was built upon ``words" (i.e. bio-words) of the protein language, in which BPE was utilized to build the bio-word vocabulary. \citet{wang2019high} suggested that BPE-segmented words indicate a language-like behavior for the protein sequences and reported improved accuracy results compared to using $3$-mers as words. 

\paragraph{Pattern-based words}  Subsequences that are conserved throughout evolution are usually associated with protein structure and function. These conserved sequences can be detected as patterns via multiple sequence alignment (MSA) techniques and Hidden Markov Models (HMM). PROSITE \cite{hulo2006prosite}, a public database that provides information on domains and motifs of proteins, uses regular expressions (i.e. RE or regex) to match these subsequences. 

Protein domains have been investigated for their potential of being the words of the protein language. One earlier study suggested that folded domains could be considered as ``phrases/clauses" rather than ``words" because of the higher semantic complexity between them \cite{gimona2006protein}. Later, domains were described as the words, and domain architectures as sentences of the language \cite{scaiewicz2015language, yu2019grammar}. 
Protein domains were treated as the words of multi-domain proteins in order to evaluate the semantic meaning behind the domains \cite{buchan2019inferring}. The study supported  prior work by \citet{yu2019grammar} suggesting that domains displayed syntactic and semantic features, but there are only a few multi-domain proteins with more than six domains limiting the use of domains as words to build sentences. Protein domains and motifs have also been utilized as words in different drug discovery tasks such as the prediction of drug-target interaction affinity \cite{greenside2017prediction, ozturk2019widedta}. These studies showed that motifs and domains together contribute to the prediction as much as the use of the full protein sequence. 

SMARTS is a well-known regex-based querying language that is used to identify patterns in a SMILES string. SMARTS has been utilized to build specific rules for small-molecule protonation \cite{ropp2019dimorphite}, to design novel ligands based on the fragments connected to the active site of a target \cite{cheron2015opengrowth}, and to help generate products in reaction prediction \cite{wei2016neural}. MolBlocks, a molecular fragmentation tool, also adopted SMARTS dictionaries to partition a SMILES string into overlapping fragments \cite{ghersi2014molblocks}. Furthermore, MACCS \cite{durant2002reoptimization} and PubChem \cite{bolton2008pubchem} Fingerprints (FP) are molecular descriptors that are described as binary vectors based on the absence/presence of substructures that are predefined with SMARTS language. A recent study on protein family clustering uses a ligand-centric representation to describe proteins in which ligands were represented with SMILES-based (i.e. 8-mers) representation, MACCS and Extended Connectivity Fingerprint (ECFP6) \cite{ozturk2018novel}. The results indicate that three of the ligand representation approaches provide similar performances for protein family clustering.

To the best of our knowledge, there is no comprehensive evaluation of the different word extraction techniques except a comparison by \citet{wang2019high} of the performance of BPE-based words against $k$-mers in a PPI prediction task. Such comparison would provide important insights to the bio/cheminformatics community.  

\subsection{Text representation}

The representation of a text (e.g. molecule or protein sequence) aims to capture syntactic, semantic or relational meaning. In the widely used Vector Space Model (VSM), a text is represented by a feature vector of either weighted or un-weighted terms \cite{salton1975vector}. The terms of this vector may correspond to words, phrases, k-grams, characters, or dimensions in a semantic space such as in the distributed word embedding representation models. The similarity between two texts represented in the vector space model is usually computed using the cosine similarity metric \cite{bilenko2003adaptive}, which corresponds to the cosine of the angle between the two vectors. 

Similarly to the one-hot encoding scheme \cite{bishop2006pattern}, in the traditional bag-of-words \cite{turney2010frequency}  and term frequency-inverse document frequency (TF-IDF) \cite{jones2004statistical}  text representation models, each word corresponds to a different dimension in the vector space. Therefore, the similarity between two words in the vector space is zero, even if they are synonymous or related to each other. In the distributed representation models \cite{mikolov2013distributed} on the other hand, words are represented as dense vectors based on their context. Words that occur in similar contexts have similar vector representations.  In this subsection, we review these commonly used text representation models with their applications in cheminformatics.  
 

\paragraph{Bag-of-words representation} In this representation model, a text is represented as a vector of \textit{bag-of-words}, where the multiplicity of the words is taken into account, but the order of the words in the text is lost \cite{turney2010frequency}. For instance, the SMILES of ampicillin ``CC1(C(N2C(S1)C(C2=O)NC(=O)C(\\C3=CC=CC=C3)N)C(=O)O)C" can be represented as a bag-of $8$-mers as follows: \{``CC1(C(N2", ``C1(C(N2C", ``1(C(N2C(", ``(C(N2C(S",...,``N)C(=O)O" ,``)C(=O)O)" ,``C(=O)O)C" \}.  We can vectorize it as $S = [1, 1, 1, 1, ...,1, 1, 1]$ in which each number refers to the frequency of the corresponding $8$-mer.


Bag-of-words representation was used in  molecular similarity computation, in which the SMILES string and the LINGOs extracted from it were treated as the \textit{sentence} and \textit{words}, respectively \cite{vidal2005}. The unique LINGOs were considered for each pair and a Tanimoto coefficient was used to measure the similarity \cite{vidal2005}. Another approach called SMILES Fingerprint (SMIfp) also adopted bag-of-words to create representations of molecules for a ligand-based virtual screening task \cite{schwartz2013smifp}. SMIfp considered 34 unique symbols in SMILES strings to create a frequency-based vector representation, which was utilized to compute molecular similarity. SMIfp provided comparable results to a chemical representation technique that also incorporated polar group and topological information, as well as atom and bond information, in recovering active compounds amongst decoys \cite{schwartz2013smifp}. 

\paragraph{TF-IDF}
The bag-of-words model, which is based on counting the terms of the sentence/document, might prioritize insignificant but frequent words. To overcome this issue, a weighting scheme can be integrated into the vector representation in order to give more importance to the rare terms that might play a key role in detecting similarity between two documents. One popular weighting approach is to use term frequency-inverse document frequency (TF-IDF) \cite{jones2004statistical}. TF refers to the frequency of a term in the document, and IDF denotes the logarithm of the total number of documents over the number of documents in which the term appears.  IDF is therefore an indicator of uniqueness.  For instance, the IDF of ``C3=CC=CC" is lower than that of ``(C(N2C(S", which appears in fewer compounds. Therefore, the existence of ``(C(N2C(S" in a compound may be more informative.  

TF-IDF weigthing was utilized to assign weights to LINGOs that were extracted from SMILES in order to compute molecule similarity using cosine similarity \cite{ozturk2016comparative}. Molecular similarities were then used as input for drug-target interaction prediction. A similar performance between TF-IDF weighted LINGO and a graph-based chemical similarity measurement was obtained. \citet{cadeddu2014organic}  used TF-IDF weighting on chemical bonds to show that bonds with higher TF-IDF scores have a higher probability of breaking. 


\paragraph{One-hot representation} 
In one-hot representation, for a given vocabulary of a text, each unique word/character is represented with a binary vector that has a $1$ in the corresponding position, while the vector positions for the remaining words/characters are filled with $0$s \cite{bishop2006pattern}. One-hot encoding is fast to build, but might lead to sparse vectors with large dimensions based on the size of the vocabulary (e.g. one million unique words in the vocabulary means one million dimensional binary vectors filled with zeros except one). It is a popular choice, especially in  machine learning-based bio/cheminformatic studies to encode different types of information such as SMILES characters \cite{segler2017generating, kwon2017deepcci},  atom/bond types \cite{preuer2019interpretable, de2018molgan} and molecular properties \cite{mayr2016deeptox}.

\paragraph{Distributed representations} 
The one-hot encoding builds discrete representations, and thus does not consider the relationships between words. For instance, the cosine similarity of two different words is 0 even if they are semantically similar. However, if the word  (i.e. $8$-mer) ``(C(N2C(S" frequently appears together with the word ``C(C2=O)N" in SMILES strings, this might suggest that they have related ``meanings". Furthermore, two words might have similar semantic meanings even though they are syntactically apart. This is where distributed vector representations come into play. 

The distributed word embeddings models gained popularity with the introduction of  Word2Vec \cite{mikolov2013distributed} and GloVe \cite{pennington2014glove}. The main motivation behind the Word2Vec model is to build real-valued high-dimensional vectors for each word in the vocabulary based on the context in which they appear. There are two main approaches in Word2Vec: (i) Skip-Gram and (ii) Continuous Bag of Words (CBOW). The aim of the Skip-Gram model is to predict context words given the center word, whereas in CBOW the objective is to predict the target word given the context words. Figure \ref{fig:word2vec} depicts the Skip-gram architecture in Word2Vec \cite{mikolov2013distributed}. For the vocabulary of size $V$, given the target word ``2C(S", the model learns to predict two context words. Both target word and context words are represented as one-hot encoded binary vectors of size $V$.  The number of neurons in the hidden layer determines the size of the embedding vectors. The weight matrix between the input layer and the hidden layer stores the embeddings of the vocabulary words. The $i^{th}$ row of the embedding matrix corresponds to the embedding of the $i^{th}$ word. 


The Word2Vec architecture has inspired a great deal of research in the bio/cheminformatics domains. The Word2Vec algorithm has been successfully applied for determining protein classes \cite{asgari2015continuous} and  protein-protein interactions (PPI) \cite{wang2019high}.  \citet{asgari2015continuous} treated $3$-mers as the words of the protein sequence and observed that $3$-mers with similar  biophysical and biochemical properties clustered together when their embeddings were mapped onto the 2D space.  \citet{wang2019high}, on the other hand, utilized BPE-based word segmentation (i.e. bio-words) to determine the words. The authors argued that the improved performance for bio-words in the PPI prediction task might be due to the segmentation-based model providing more distinct words than $k$-mers, which include repetitive segments. Another recent study treated  multi-domain proteins as sentences in which each domain was recognized as a word \cite{buchan2019inferring}. The Word2Vec algorithm was trained on the domains (i.e. PFAM domain identifiers) of eukaryotic protein sequences to learn semantically interpretable representations of them. The domain representations were then investigated in terms of the Gene Ontology (GO) annotations that they inherit. The results indicated that semantically similar domains share similar GO terms. 

The Word2Vec algorithm was also utilized for representation of chemicals. SMILESVec, a text-based ligand representation technique, utilized Word2Vec to learn embeddings for $8$-mers (i.e. chemical words) that are extracted from SMILES strings \cite{ozturk2018novel}. SMILESVec was utilized in protein representation such that proteins were represented as the average of the SMILESVec vectors  of their interacting ligands. The results indicated comparable performances for ligand-based and sequence based protein representations in protein family/superfamily clustering. Mol2Vec \cite{jaeger2018mol2vec}, on the other hand, was based on the identifiers of the substructures (i.e. words of the chemical) that were extracted via Extended Connectivity Fingerprint (ECFP) \cite{rogers2010extended}. The results showed a better performance with Mol2Vec than with the simple Morgan Fingerprint in a solubility prediction task, and a comparable performance to graph-based chemical representation \cite{wu2018moleculenet}. \citet{chakravarti2018distributed} also employed the Word2vec model that was trained on the fragments that are extracted from SMILES strings using a graph traversing algorithm.  The results favored the distributed fragment-based ligand representation over fragment-based binary vector representation in a ring system clustering task and showed a comparable performance in the prediction of toxicity against \textit{Tetrahymena} \cite{chakravarti2018distributed}. Figure \ref{fig:scheme} illustrates the pipeline of a text-based molecule representation based on $k$-mers. 


FP2Vec is another method that utilizes embedding representation for molecules, however instead of the Word2Vec algorithm, it depends on a Convolutional Neural Network (CNN) to build molecule representations to be used in toxicity prediction tasks \cite{jeon2019fp2vec}.  CNN architectures have also been utilized for drug-target binding affinity prediction \cite{ozturk2018deepdta} and drug-drug interaction prediction \cite{kwon2017deepcci} to build representations for chemicals from raw SMILES strings, as well as for protein fold prediction \cite{hou2017deepsf} to learn  representations for proteins from amino-acid sequences.  SMILES2Vec adopted different DL architectures (GRU, LSTM, CNN+GRU, and CNN+LSTM) to learn molecule embeddings, which were then used to predict toxicity, affinity and solubility \cite{goh2017smiles2vec}. A CNN+GRU combination was better at the prediction of chemical properties. A recent study compared several DL approaches to investigate the effect of different chemical representations, which were learned through these architectures, on a chemical property prediction problem \cite{paul2018chemixnet}. The authors also combined DL architectures that were trained on SMILES strings with the MACCS fingerprint, proposing a combined representation for molecules (i.e. CheMixNet). The CheMixNet representation outperformed the other representations that were trained on a single data type such as SMILES2Vec (i.e. SMILES) and Chemception (i.e. 2D graph) \cite{goh2017chemception}.

\subsection{Text generation} \label{section:textgeneration}

Text generation is a primary NLP task, where the aim is to generate grammatically and semantically correct text, with many applications ranging from question answering to machine translation \cite{wang2019topic}. It is generally formulated as a language modeling task, where a statistical model is trained using a large corpus to predict the distribution of the next word in a given context. In machine translation, the generated text is the translation of an input text in another language. 

Medicinal chemistry campaigns use methods such as scaffold hopping \cite{grisoni2018scaffold} or fragment-based drug design \cite{sledz2018protein} to build and test novel molecules but the chemotype diversity and novelty may be limited. It is possible to explore uncharted chemical space with text generation models, which learn a distribution from the available data  (i.e. SMILES language) and generate novel molecules that share similar physicochemical properties with the existing molecules \cite{segler2017generating}. Molecule generation can then be followed by assessing physicochemical properties of the generated compound or its binding potential to a target protein \cite{segler2017generating}. For a comprehensive review of molecule generation methodologies, including graph-based models, we refer the reader to the review of \citet{elton2019deep}. Machine translation models have also been recently adapted to text-based molecule generation, which start with one ``language" such as that of reactants and generate a novel text in another ``language" such as that of products \cite{schwaller2018molecular}. Below, we present recent studies on text based molecule generation.

RNN models, which learn a probability distribution from a training set of molecules, are commonly used in molecule generation to propose novel molecules similar to the ones in the training data set. For instance, given the SMILES sequence ``C(=O", the model would predict the next character to be ``)" with a higher probability than ``(". The production of valid SMILES strings, however, is a challenge because of the complicated SMILES syntax that utilizes parentheses to indicate branches and ring numbers. The sequential nature of RNNs, which may miss long range dependencies, is a disadvantage of these models \cite{segler2017generating}. RNN descendants LSTM and GRU, which model long-term dependencies, are better suited for remembering matching rings and branch closures. Motivated by such a hypothesis,  \citet{segler2017generating} and \citet{ertl2017silico} successfully pioneered \textit{de novo} molecule generation using LSTM architecture to generate valid novel SMILES. \citet{segler2017generating} further modified their model to generate target-specific molecules by integrating a target bioactivity prediction step to filter out inactive molecules and then retraining the LSTM network.  In another study, transfer learning was adopted to fine-tune an LSTM-based SMILES generation model so that structurally similar leads were generated for targets with few known ligands  \cite{gupta2018generative}. \citet{olivecrona2017molecular} and \citet{popova2018deep} used reinforcement learning (RL) to bias their model toward compounds with desired properties. Merk et al. \cite{merk2018novo, merk2018tuning} fine-tuned their LSTM model on a target-focused library of active molecules and synthesized some novel compounds. \citet{arus2019exploring} explored how much of the GDB-13 database \cite{blum2009970} they could rediscover by using an RNN-based generative model. 

The variational Auto-encoder (VAE) is another widely adopted text generation architecture \cite{bowman2015generating}. \citet{gomez2018automatic} adopted this architecture for molecule generation. A traditional auto-encoder encodes the input into the latent space, which is then decoded to reconstruct the input. VAE differs from AE by explicitly defining a probability distribution on the latent space to generate new samples. \citet{gomez2018automatic}  hypothesized that the variational part of the system integrates noise to the encoder, so that the decoder can be more robust to the large diversity of molecules. However, the authors also reported that the non-context free property of SMILES caused by matching ring numbers and parentheses might often lead the decoder to generate invalid SMILES strings.  A grammar variational auto-encoder (GVAE), where the grammar for SMILES is explicitly defined instead of the auto-encoder learning the grammar itself, was proposed to address this issue \cite{kusner2017grammar}. This way, the generation is based on the pre-defined grammar rules and the decoding process generates grammar production rules that should also be grammatically valid. Although syntactic validity would be ensured, the molecules may not have semantic validity (chemical validity). \citet{dai2018syntax} built upon the VAE \cite{gomez2018automatic} and GVAE \cite{kusner2017grammar} architectures and introduced a syntax-directed variational autoencoder (SD-VAE) model for the molecular generation task. The syntax-direct generative mechanism in the decoder contributed to creating  both syntactically and semantically valid SMILES sequences. \citet{dai2018syntax} compared the latent representations of molecules generated by VAE, GVAE, and SD-VAE, and showed that SD-VAE provided better discriminative features for druglikeness. \citet{blaschke2018application} proposed an adversarial AE for the same task. Conditional VAEs \cite{lim2018molecular, kang2018conditional} were trained to generate molecules conditioned on a desired property. The  challenges that SMILES syntax presents inspired the introduction of new syntax such as DeepSMILES \cite{OBoyle2018} and SELFIES \cite{krenn2019selfies} (details in Section \ref{section:availabledata}). 

Generative Adversarial Network (GAN) models generate novel molecules by using two components: the generator network generates novel molecules, and the discriminator network aims to distinguish between the generated molecules and real molecules \cite{hong2019generative}. In text generation models, the novel molecules are drawn from a distribution, which are then fine-tuned to obtain specific features, whereas adversarial learning utilizes generator and discriminator networks to produce novel molecules \cite{hong2019generative, guimaraes2017objective}. ORGAN \cite{guimaraes2017objective}, a molecular generation methodology, was built upon a sequence generative adversarial network (SeqGAN) from NLP \cite{yu2017seqgan}. ORGAN integrated RL in order to generate molecules with desirable properties such as solubility, druglikeness, and synthetizability through using domain-specific rewards  \cite{guimaraes2017objective}.

\paragraph{Machine Translation} Machine translation finds use in cheminformatics in ``translation" from one language (e.g. reactants) to another (e.g. products). Machine translation is a challenging task because the syntactic and semantic dependencies of each language differ from one another and this may give rise to ambiguities. Neural Machine Translation (NMT) models benefit from the potential of deep learning architectures to build a statistical model that aims to find the most probable target sequence for an input sequence by learning from a corpus of examples \cite{sutskever2014sequence, cho2014learning}. The main advantage of NMT models is that they provide an end-to-end system that utilizes a single neural network to convert the source sequence into the target sequence. \citet{sutskever2014sequence} refer to their model as a sequence-to-sequence (seq2seq) system that addresses a major limitation of DNNs that can only work with fixed-dimensionality information as input and output. However, in the machine translation task, the length of the input sequences is not fixed, and the length of the output sequences is not known in advance. 

The NMT models are based on an encoder-decoder architecture that aims to maximize the probability of generating the target sequence (i.e. most likely correct translation) for the given source sequence. The first encoder-decoder architectures in NMT performed poorly as the sequence length increased mainly because the encoder mapped the source sequence into a single fixed-length vector. However, fixed-size representation may be too small to encode all the information required to translate long sequences \cite{bahdanau2014neural}. To overcome the issue of the fixed context vector (Figure  \ref{fig:smilestranslate}a), a new method was developed, in which every source token was encoded into a memory bank independently (Figure \ref{fig:smilestranslate}b). The decoder could then selectively focus on parts of this memory bank during translation \cite{bahdanau2014neural, luong2015effective}. This technique is known as ``attention mechanism" \cite{graves2013generating}. 


Inspired by the successes in NMT, the first application of seq2seq models in cheminformatics was for reaction prediction by  \citet{nam2016linking}, who proposed to translate the SMILES strings of reactants and separated reagents to the corresponding product SMILES. The authors hypothesized that the reaction prediction problem can be re-modelled as a translation system in which both inputs and output are sequences.  Their model used GRUs for the encoder-decoder and a Bahdanau \cite{bahdanau2014neural} attention layer in between. \citet{liu2017retrosynthetic} in contrast, performed the opposite task, the single-step retrosynthesis prediction, using a similar encoder-decoder model. When given a product and a reaction class, their model predicted the reactants that would react together to form that product. One major challenge in the retrosynthesis prediction task is the possibility of multiple correct targets, because more than one reactant combination could lead to the same product. Similarly to \citet{nam2016linking}, \citet{schwaller2018found} also adopted a seq2seq model to translate precursors into products, utilizing the SMILES representation for the reaction prediction problem. Their model used a different attention mechanism by \citet{luong2015effective}  and LSTMs in the encoder and decoder. By visualizing the attention weights, an atom-wise mapping between the product and the reactants could be obtained and used to understand the predictions better. \citet{schwaller2018found} showed that seq2seq models could compete with graph neural network-based models in the reaction prediction task \cite{jin2017predicting}. 

A translation model was also employed to learn a data-driven representation of molecules \cite{winter2019learning}. \citet{winter2019learning} translated between two textual representations of a chemical, InChi and SMILES, to extract latent representations that can integrate the semantic ``meaning" of the molecule.  The results indicated a statistically significant improvement with the latent representations in a ligand-based virtual screening task against fingerprint methods such as ECFP (i.e. Morgan algorithm). NMT architectures were also adopted in a protein function prediction task for the first time, in which ``words" that were extracted from protein sequences are translated into GO identifiers using RNNs as encoder and decoder \cite{cao2017prolango}. Although exhibiting a comparable performance to the state-of-the-art protein function prediction methods, the authors argued that the performance of the model could be improved by determining more meaningful ``words" such as biologically interpretable fragments.

Transformer is an attention-based encoder-decoder architecture that was introduced in NMT by \citet{Vaswani:2017ul}. Although similar to previous studies \cite{sutskever2014sequence, cho2014learning,  bahdanau2014neural} in terms of adopting an encoder-decoder architecture, Transformer differs from the others because it only consists of attention and feed-forward layers in the encoder and decoder.  As transformers do not contain an RNN,  positional embeddings are needed to capture order relationships in the sequences. \citet{schwaller2018molecular} were the first to adopt the Transformer architecture in cheminformatics and designed a \textit{Molecular Transformer} for the chemical reaction prediction task. The Molecular Transformer, which was atom-mapping independent, outperformed the other algorithms (e.g. based on a two-step convolutional graph neural network \cite{coley2019graph}) on commonly used benchmark data sets. Transformer architecture was also adopted to learn representations for chemicals in prediction of drug-target interactions \cite{shin2019self} and molecular properties \cite{wang2019smiles} in which the proposed systems either outperformed the state-of-the-art systems or obtained comparable results.

\section{Future Perspectives}
The increase in the biochemical data available in  public databases combined with the advances in computational power and NLP methodologies have given rise to a rapid growth in the publication rate in bio/cheminformatics, especially through pre-print servers. As this interdisciplinary field grows, novel opportunities come hand in hand with novel challenges. 

\subsection{Challenges}
The major challenges that can be observed from investigating these studies can be summarized as follows: (i) the need for universalized benchmarks and metrics, (ii) reproducibility of the published methodologies, (iii) bias in available data, and (iv) biological and chemical interpretability/explainability of the solutions. 

\paragraph{Benchmarking} There are several steps in the drug discovery pipeline, from affinity prediction to the prediction of other chemical properties such as toxicity, and solubility. The use of different datasets and different evaluation metrics makes the assessment of model performance challenging.  Comprehensive benchmarking platforms that can assess the success of different tools are still lacking. A benchmarking environment rigorously brings together the suitable data sets and evaluation methodologies in order to provide a fair comparison between the available tools. Such environments are available for molecule generation task from MOSES \cite{polykovskiy2018molecular} and GuacaMol \cite{brown2019guacamol}. \textit{MoleculeNet} is also a similar attempt to build a benchmarking platform for tasks such as prediction of binding affinity and toxicity \cite{wu2018moleculenet}.

\paragraph{Reproducibility} Despite the focus on sharing datasets and source codes on popular software development platforms such as GitHub (github.com) or Zenodo (zenodo.org), it is still a challenge to use data or code from other groups. The use of FAIR (Findable, Accessible, Interoperable and Reusable) (meta)data principles can guide the management of scientific data \cite{wilkinson2016fair}. Automated workflows that are easy to use and do not require programming knowledge encourage the flow of information from one discipline to the other. Platform-free solutions such as Docker (docker.com) in which an image of the source code is saved and can be opened without requiring further installation could accelerate the reproduction process. A recent initiative to provide a unified-framework for predictive models in genomics can quickly be adopted by the medicinal chemistry community \cite{avsec2019kipoi}. 

\paragraph{Bias in data} The available data has two significant sources of bias, one related to the limited sampling of chemical space and the other related to the quality and reproducibility of the data. The lack of information about some regions of the protein/chemical landscape limits the current methodologies to the exploitation of data rather than full exploration. The data on  protein-compound interactions is biased toward some privileged molecules or proteins because the protein targets  are related to common diseases or the molecules  are similar to known actives. Hence, not all of chemical space is sampled, and  chemical space is expanded based on the similarity of an active compound to others, which is also referred to as  inductive bias \cite{cleves2008effects}. Data about proteins or molecules related to rare diseases is limited and inactive molecules are frequently not reported.  Moreover, some experimental measurements that are not reproducible across different labs or conditions limit their reliability \cite{pogue2018rare}. \citet{sieg2019need} and \citet{zhang2019bayesian} have recently discussed the bias factors in dataset composition. Zhang and Lee have also addressed the sources of bias in the data and proposed to use Bayesian deep learning to quantify uncertainty.

\paragraph{Interpretability} The black box nature of ML/DL methodologies makes assigning meaning to the results difficult. Explainability of an ML model is especially critical in drug discovery to facilitate the use of these findings by medicinal chemists, who can contribute to the knowledge loop. \textit{explainable-AI} (XAI) is a current challenge that calls for increased interpretability of  AI solutions for a given context and includes several factors such as trust, safety, privacy, security, fairness and confidence \cite{holzinger2017we}. Explainability is also critical for the domain experts to assess the reliability of new methodolodogies. Interpretability is usually classified into two categories: post-hoc (i.e. after) and ante-hoc (i.e. before). Post-hoc approaches explain the predictions of the model, whereas ante-hoc approaches integrate explainability into the model. Recent studies have already aimed to map the semantic meaning behind the models onto the biochemical description. An attentive pooling network, a two-way attention system that extends the attention mechanism by allowing input nodes to be aware of one another, is one approach that has been employed in drug-target interaction prediction \cite{gao2018interpretable}. \citet{preuer2019interpretable} showed that mapping activations of hidden neurons in feed-forward neural networks to pharmacophores, or linking atom representations computed by convolutional filters to substructures in a graph-convolution model,  are possible ways of integrating  explainability into  AI-based drug discovery systems.
\citet{bradshaw2019} also demonstrated a novel approach that combines molecule generation and retrosynthesis prediction to generate synthesizable molecules. Integration of such solutions to drug discovery problems will not only be useful for computational researchers but also for the medicinal chemistry community.

\subsection{Opportunities}

The NLP field has seen tremendous advances in the past five years, starting with the introduction of distributed word embedding algorithms such as Word2Vec \cite{mikolov2013distributed} and Glove \cite{pennington2014glove}.  The concept of contextualized word embeddings (i.e. ELMo) was introduced soon after \cite{peters2018deep}. Here, the embedding of the word is not fixed, but changes according to the context (i.e. sentence) in which it appears. These advances continued with more complicated architectures such as  Transformer (i.e. Generative Pre-Training or GPT) \cite{radford2018improving}  and BERT \cite{devlin2018bert}, RoBERTa \cite{liu2019roberta}, GPT2 \cite{radford2019language}, Transformer-XL \cite{dai2019transformer}, and XLNet \cite{yang2019xlnet} models. Such models with a focus on context might have significant impact not only on drug discovery, but also on the protein folding problem, which is critical for predicting structural properties of the protein partner.  Secondary structure \cite{hanson2019getting, zhu2019predicting, wang2016protein}, domain boundary \cite{shi2019dnn} and fold \cite{wei2015enhanced} prediction studies often use sequence information in combination with similarity to available structures.  The recent success of AlphaFold \cite{evans2018novo} in Critical Assessment of Protein Structure Prediction (CASP) competitions (\url{http://predictioncenter.org/}) showed that the enhanced definitions of context, brought about by the advances in machine/deep learning systems, might be useful for capturing the global dependencies in protein sequences to detect interactions between residues separated in sequence space but close together in 3D space \cite{hanson2019getting}.

Unsupervised learning can be used on ``big" textual data through using language models with attention \cite{Vaswani:2017ul} and using pre-trained checkpoints from language models \cite{Rothe:2019wo}. Encoder-decoder architectures have also had significant impact on solving text generation and machine translation problems and were successfully applied to molecule generation problem. As NLP moves forward, the most recent approaches such as Topic-Guided VAE \cite{wang2019topic} and knowledge graphs with graph transformers \cite{koncel2019text} will easily find application in bio/cheminformatics.

Recent NLP models are not domain-specific, and they can help with the generalization of  models  \cite{radford2019language}. Current studies emphasize multi-task learning, which requires the use of DNNs that share parameters to learn more information from related but individual tasks \cite{ruder2019neural, radford2019language}. Combined with the transferability of contextual word representation models, multi-task learning can also provide solutions to drug discovery which has many interwoven tasks, such as chemical property prediction and molecule generation.

Language has an important power, not only for daily communication but also for the communication of codified domain knowledge. Deciphering the meaning behind text is the primary purpose of NLP, which inevitably has found its way to bio/cheminformatics. The complicated nature of  biochemical text makes understanding the semantic construction of the hidden words all the more challenging and interesting. The applications we discussed in this review provide a broad perspective of how NLP is already integrated with the processing of biochemical text. A common theme in all of these applications is the use of AI-based methodologies that drive and benefit from the NLP field. Novel advances in NLP and ML are providing auspicious results to solving long-standing bio/cheminformatics problems.  

With this review, we have summarized the impact of NLP on bio/cheminformatics to encourage this already interdisciplinary field to take advantage of recent advances. The communication between researchers from different backgrounds and domains can be enhanced through establishing a common vocabulary toward common goals. This review has been an attempt to facilitate this conversation.

\section*{Acknowledgement}
This work is partially supported by TUBITAK (The Scientific and Technological Research Council of Turkey) under grant number 119E133. HO acknowledges TUBITAK-BIDEB 2211 scholarship program and thanks G\"{o}k\c{c}e Uludo\u{g}an for her comments on figures. EO thanks Prof. Amedeo Caflisch for hosting her at the University of Zurich during her sabbatical.



\bibliography{elsarticle-template}

\begin{figure}[H]%
\centerline{\includegraphics[scale=0.45]{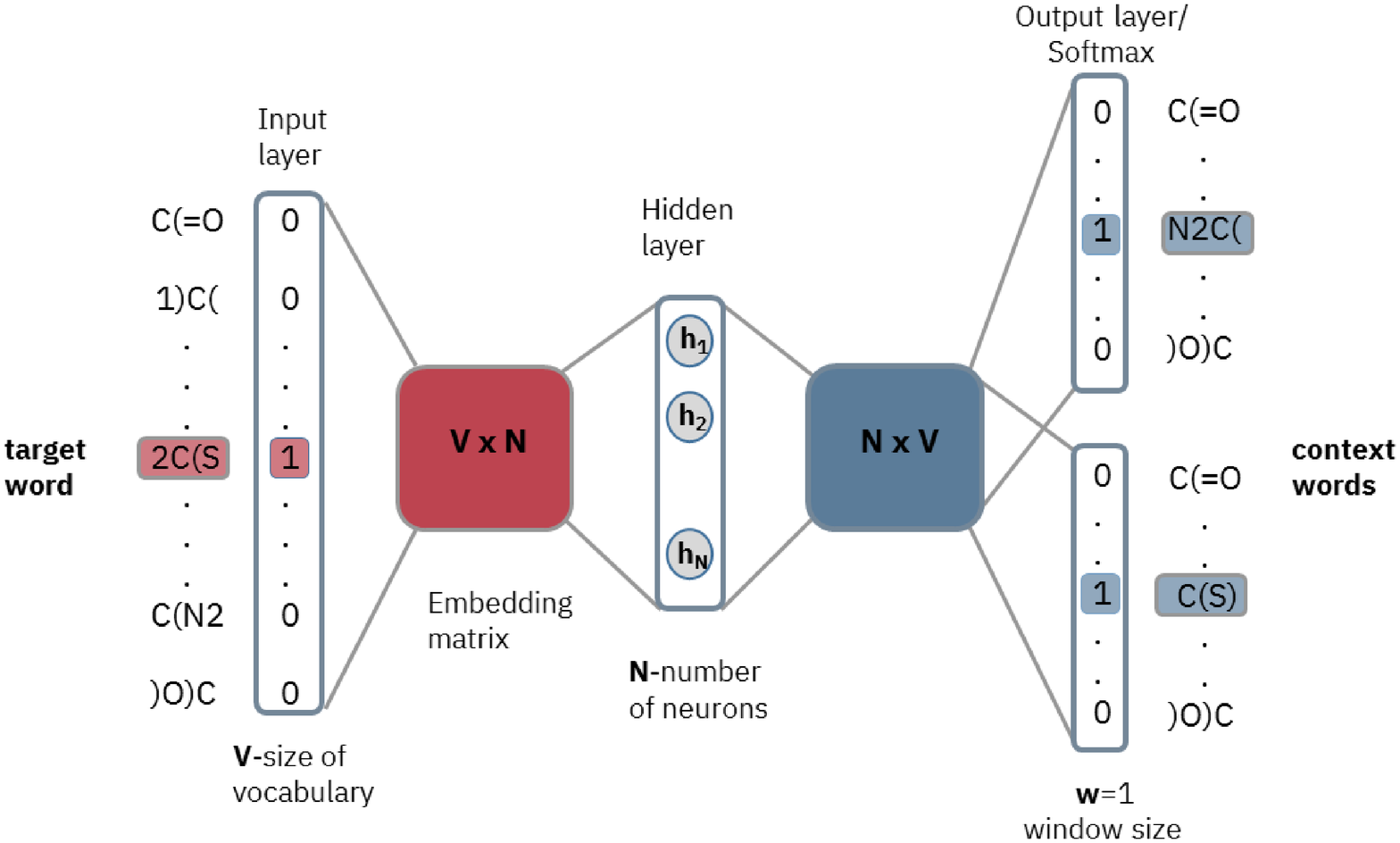}} 
\caption{ The illustration of the Skip-Gram architecture of the Word2Vec algorithm. For a vocabulary of size V, each word in the vocabulary is described as a one-hot encoded vector (a binary vector in which only the corresponding word position is set to 1). The Skip-Gram architecture is a simple one hidden-layer neural network that aims to predict context (neighbor) words of a given target word. The extent of the context is determined by the window size parameter. In this example, the window size is equal to 1, indicating that the system will predict two context words (the word on the left and the word on the right of the target word) based on their probability scores. The number of nodes in the hidden layer (N) controls the size of the embedding vector. The weight matrix of VxN stores the trained embedding vectors.  }\label{fig:word2vec}
\end{figure}

\begin{figure}[H]%
\centerline{\includegraphics[scale=0.45]{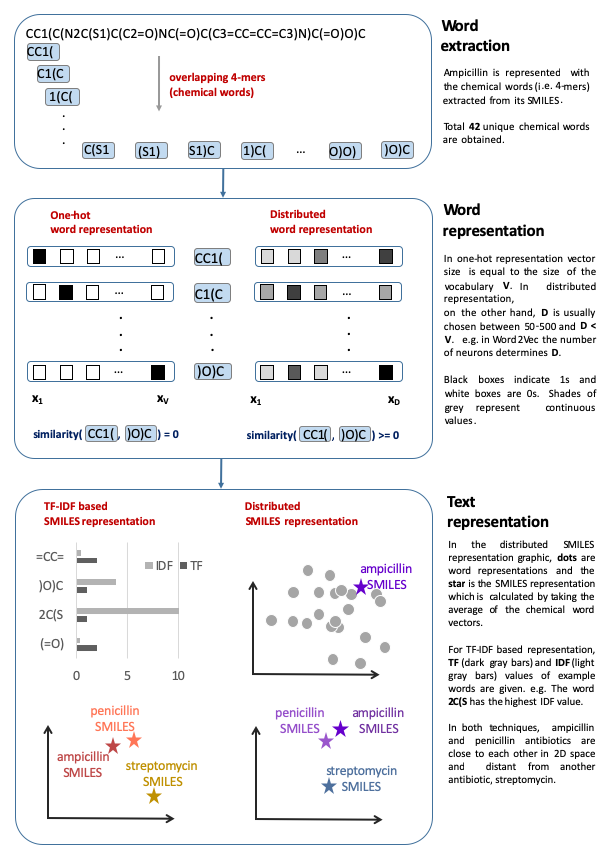}} 
\caption{ (Continued on the following page.) }\label{fig:scheme}
\end{figure}
\begin{figure}[H]
  \contcaption{The workflow for building a SMILES-based molecule representation.  In the first box, SMILES text of ampicillin is utilized to extract words. In this case, the words are overlapping 4-mers and there are total 42 unique words. To represent multiple compounds, words are extracted from each compound, thus building a vocabulary of size $V$. In the second box, two popular word representations are illustrated: (left) one-hot encoded representation, and (right) distributed representation. With the one-hot encoding, we build a binary vector of size $V$, in which the position of the corresponding word is set to 1, while the rest remains as 0. In the distributed representations, however, the dimension of the word representation (embedding) is $D$, which is usually smaller than $V$ and $50 < D < 500$. Furthermore, distributed representations are continuous vectors. Therefore, the cosine similarity of two distributed word vectors is equal to or greater than 0, whereas with one-hot encoded word vectors, their similarity is 0 if they are not equal. Finally, the third box demonstrates the text level representation. The analogy between texts and SMILES strings allows us to represent chemicals as groups of ``chemical words". Term-Frequency-Inverse Document Frequency (TF-IDF) weighting, which is a widely adopted weighting scheme in Information Retrieval domain, assigns higher weights to rare words. In a corpus with a vocabulary size of V, each word is represented as the multiplication of its frequency and IDF values. In the distributed representation of texts, since each word also has a D dimensional embedding vector, text representation is computed based on these word embedding vectors, for example by taking their average. Dots (.) in the third box represent ``chemical words" in 2D space, whereas stars (*) represent the whole SMILES (i.e. compound). With both techniques, the compound (i.e. text)  representations can be mapped to 2D. We expect chemicals such as ampicillin and penicillin, which are from the same antibiotic class, to be close to each other in vector space, whereas streptomycin, an antibiotic from a different class, to be distant. }
\end{figure}

\begin{figure}[H]%
\centerline{\includegraphics[scale=0.6]{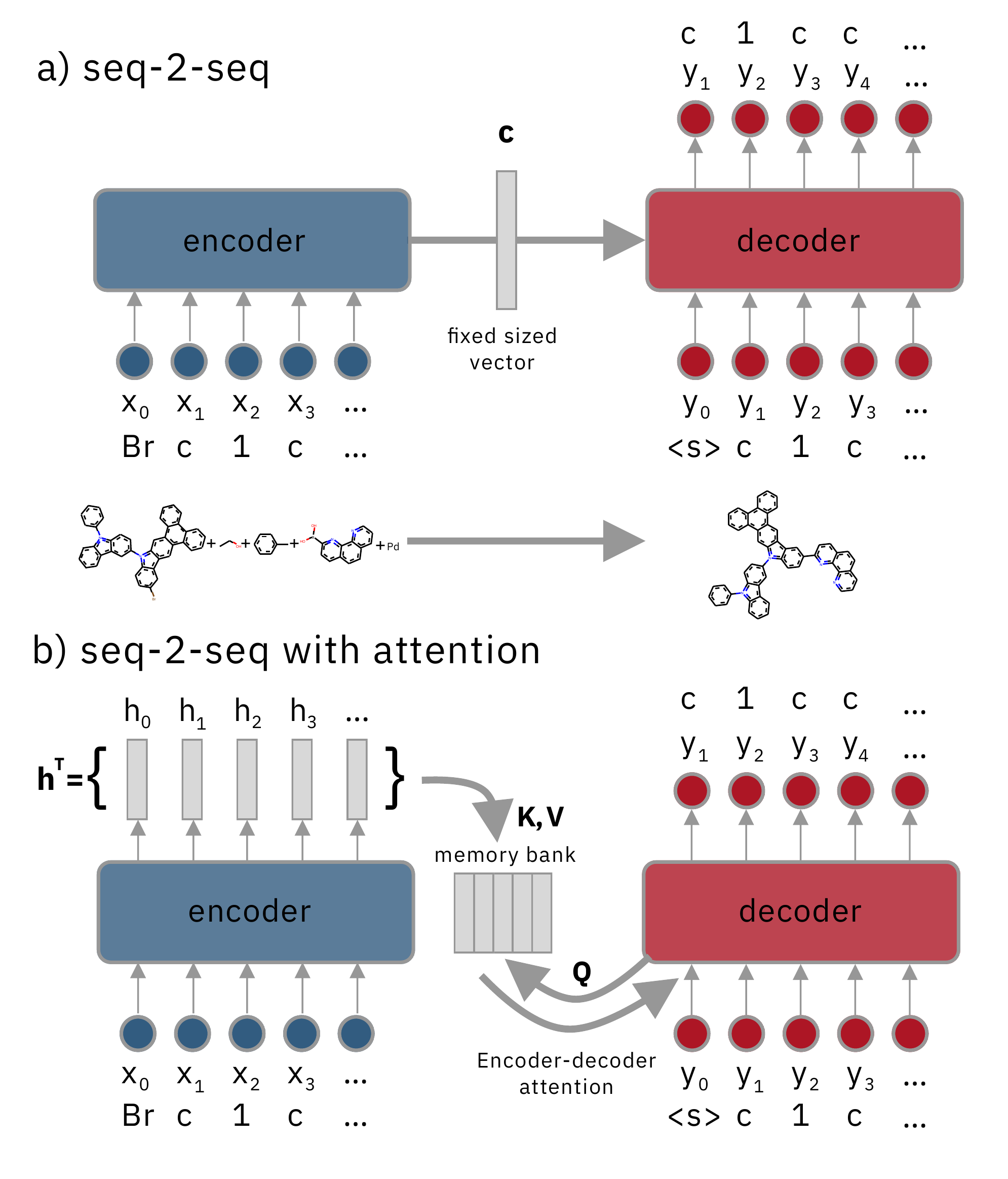}} 
\caption{(Continued on the following page.) 
 }\label{fig:smilestranslate}
\end{figure}
\begin{figure}[H]
  \contcaption{Sequence-2-Sequence models take as input a sequence of tokens and generate a sequence of tokens as output. The example in this Figure is a chemical reaction prediction, where given a set of precursors the most likely products are predicted.  The input tokens correspond to the tokenized SMILES of the precursors and the generated tokens to the SMILES of the product. In the original sequence-2-sequence models, the encoder encoded the input sequence into a fixed size context vector, as shown in (a). The decoder had access only to this fixed size vector, which limited its application for long input sequences. To overcome this drawback, the attention mechanism was introduced, as shown in (b). In a sequence-2-sequence model with attention, the encoder encodes every token independently into a memory bank. The longer the input sequence  is, the larger is the memory bank. The decoder then queries the memory bank at every decoding step and selectively attends the most relevant value vectors to predict the next token.}
\end{figure}

\begin{table}[H]
\caption{NLP concepts and their applications in drug discovery} \label{tab:nlpconcepts}
\scalebox{0.7}{
\begin{tabular}{|l|l|l|l|} \hline
\textbf{Concept}           & \multicolumn{1}{c|}{\textbf{Definition}}    & \textbf{Methodologies}   & \textbf{Applications}      \\ \hline

\multirow{4}{*}{\textbf{Token/word}}     & \multirow{4}{*}{\begin{tabular}[c]{@{}l@{}} A series of characters (i.e. word, number, symbol) \\  that  constitutes the smallest unit of a language.  \\ The identification  of tokens (i.e. tokenization) is an \\ important  pre-processing step in many  NLP tasks, \\ e.g. substructures of a  molecule. \end{tabular}} 
& k-mers      & 
\begin{tabular}[c]{@{}l@{}}
protein family classification \cite{asgari2015continuous, ozturk2018novel} \\
protein function prediction \cite{cao2017prolango, ranjan2019deep} \\
protein language analysis \cite{motomura2012word} \\
molecular similarity \cite{vidal2005, ozturk2016comparative}
\end{tabular} \\ \cline{3-4}  &                                                

& patterns    & 
\begin{tabular}[c]{@{}l@{}}
drug-target interaction prediction\cite{greenside2017prediction, ozturk2019widedta} \\ 
protein language analysis \cite{gimona2006protein, scaiewicz2015language, yu2019grammar, buchan2019inferring} \\
molecule fragmentation \cite{ghersi2014molblocks} \\
reaction prediction \cite{wei2016neural} \\
ligand design \cite{cheron2015opengrowth}
\end{tabular}                          \\ \cline{3-4}  &                           
& MCS         & chemical language analysis \cite{cadeddu2014organic, wozniak2018linguistic}  \\ \cline{3-4}  &  

& BPE        & protein-protein interaction prediction \cite{wang2019high}  \\ \cline{3-4}  & 

& MDL  & 
protein family classification \cite{ganesan2017protein} 
                      \\ \hline

\multirow{4}{*}{\textbf{Sentence}} & 
\multirow{4}{*}{\begin{tabular}[c]{@{}l@{}} A text containing one or more tokens/words, \\ e.g. textual representations of chemicals and \\ proteins. \end{tabular}}  & 

SMILES \cite{weininger1988smiles}          & 
\begin{tabular}[c]{@{}l@{}}
molecular property prediction \cite{paul2018chemixnet} \\
binding affinity prediction \cite{ozturk2018deepdta, goh2017smiles2vec} \\
reaction prediction \cite{nam2016linking, liu2017retrosynthetic, schwaller2018found, schwaller2018molecular} \\
data augmentation \cite{bjerrum2017smiles} \\
and more.
\end{tabular}
\\ \cline{3-4}  &                 
& DeepSMILES  \cite{OBoyle2018}      &  binding affinity prediction \cite{ozturk2018chemical} \\ \cline{3-4}  &                          

& SELFIES \cite{krenn2019selfies}          & -                             \\ \cline{3-4} &                         

& protein sequence & 
\begin{tabular}[c]{@{}l@{}}
toxicity prediction \cite{jaeger2018mol2vec} \\
protein family classification \cite{asgari2015continuous, ganesan2017protein} \\
protein function prediction \cite{cao2017prolango, ranjan2019deep} \\
protein language analysis \cite{motomura2012word} \\
and more.
\end{tabular}

\\ \hline

\multirow{2}{*}{\textbf{\begin{tabular}[c]{@{}c@{}}Word/sentence\\ representation\end{tabular}}} & \multirow{2}{*}{\begin{tabular}[c]{@{}l@{}} The aim to describe a text  that can reflect its syntactic\\ and semantic features, \\ e.g. vector representation of SMILES based on \\ the occurrences of each symbol. \end{tabular}}                                                                       
& bag-of-words    & 
\begin{tabular}[c]{@{}l@{}}
molecular similarity \cite{vidal2005, ozturk2016comparative} 
\end{tabular}                        \\ \cline{3-4}  &     
& \begin{tabular}[c]{@{}l@{}}distributed\\ representations\end{tabular} &  \begin{tabular}[c]{@{}l@{}} 
binding affinity prediction \cite{ozturk2018deepdta} \\ 
chemical property prediction \cite{jaeger2018mol2vec, chakravarti2018distributed} \\
toxicity prediction \cite{goh2017smiles2vec, jaeger2018mol2vec, chakravarti2018distributed, jeon2019fp2vec} \\
drug-drug interaction prediction \cite{kwon2017deepcci} \\
protein family classification \cite{asgari2015continuous, ozturk2018novel} \\
protein-protein interaction prediction \cite{wang2019high}
\end{tabular}       \\ \hline



\multirow{2}{*}{\textbf{\begin{tabular}[c]{@{}l@{}}Machine\\ translation\end{tabular}} } & 
\multirow{2}{*}{\begin{tabular}[c]{@{}l@{}}The task of converting a sequence of meaningful\\  symbols in one language into a meaningful sequence\\ in another language,\\ e.g.translating SMILES to InChi in molecules. \end{tabular} } 

& \begin{tabular}[c]{@{}l@{}}RNN-based \\ seq2seq  \end{tabular}    & 
\begin{tabular}[c]{@{}l@{}}
protein function prediction \cite{cao2017prolango}  \\
chemical representation \cite{winter2019learning} \\
reaction prediction \cite{nam2016linking, schwaller2018found} \\
retrosynthesis \cite{liu2017retrosynthetic}
\end{tabular} 
\\ \cline{3-4}  &    
& Transformer &  \begin{tabular}[c]{@{}l@{}} 
reaction prediction \cite{schwaller2018molecular} \\
drug-target interaction prediction \cite{shin2019self} \end{tabular}  \\ \hline

\multirow{2}{*}{\textbf{\begin{tabular}[c]{@{}c@{}}Language\\ generation\end{tabular}} } & 
\multirow{2}{*}{\begin{tabular}[c]{@{}l@{}} The aim to generate a sequence of meaningful symbols\\ in the given language that are close to real.\\ e.g. generating SMILES of a novel lead\end{tabular}} & 

RNN-types          & molecule generation \cite{segler2017generating, ertl2017silico, gupta2018generative, olivecrona2017molecular, yang2017chemts}        \\ \cline{3-4}  &                          
                          
& VAE-types       &  molecule generation \cite{gomez2018automatic, kusner2017grammar, dai2018syntax}  \\ \cline{3-4}  &                     

& GAN          & molecule generation \cite{guimaraes2017objective, prykhodko2019novo}                                                          \\ \hline

\end{tabular}}
\end{table}

\begin{table}[H]
\caption{Widely used AI methodologies in NLP-based drug discovery studies} \label{tab:methodologies}
\scalebox{0.83}{
\begin{tabular}{|l|l|}
\hline
\textbf{Model}   & \textbf{Description}  \\ \hline
Deep Neural Network (DNN) \cite{bengio2009learning}  &  \begin{tabular}[c]{@{}l@{}} An artificial neural network (ANN) witha large number  \\ of hidden layers and neurons. \end{tabular} \\ \hline

Word2Vec \cite{mikolov2013distributed}  &  \begin{tabular}[c]{@{}l@{}} An ANN-based word embedding architecture that \\  captures the semantic information of the words  based  \\  on the context in which they appear.  \end{tabular} \\ \hline

Convolutional Neural Network (CNN) \cite{lecun1998gradient}  & A type of ANN that utilizes convolutions in the layers.  \\ \hline

Recurrent Neural Network (RNN) \cite{goodfellow2016deep}  &  \begin{tabular}[c]{@{}l@{}} A type of ANN that has a feedback loop connected to \\ previous time samples. \end{tabular} \\ \hline

Long-short Term Memory (LSTM) \cite{hochreiter1997long}  &  
\begin{tabular}[c]{@{}l@{}}
A type of RNN that captures long distance dependencies \\
and comprises update, forget, and output gates. 
\end{tabular}
\\ \hline

Gated Recurrent Unit   & 
\begin{tabular}[c]{@{}l@{}}
A type of RNN that captures long distance dependencies \\
and comprises an update gate.
\end{tabular}
\\ \hline

Auto-encoder (AE) \cite{goodfellow2016deep}  & \begin{tabular}[c]{@{}l@{}}  A neural network based architecture that comprises  an \\ encoder that maps the input in a narrow space and  a \\  decoder  that reconstructs the compressed representation. \end{tabular} \\ \hline

Variational Auto-encoder (VAE) \cite{kingma2013auto}  & \begin{tabular}[c]{@{}l@{}}  A type of AE that generates outputs based on a specific \\ distribution. \end{tabular}  \\ \hline

Generative Adversarial Network (GAN) \cite{guimaraes2017objective}  & \begin{tabular}[c]{@{}l@{}}  A generative model with generator and discriminator \\ networks.\end{tabular} \\ \hline

Sequence-to-sequence (seq2seq)  & \begin{tabular}[c]{@{}l@{}}  An encoder-decoder based architecture   that maps  an \\ input  sequence into an output sequence.  \end{tabular} \\ \hline

Attention mechanism \cite{bahdanau2014neural} & \begin{tabular}[c]{@{}l@{}}  enables the model to choose  among   the important  parts \\of a sequence that are relevant  to the  output.  \end{tabular} \\ \hline

Transformer \cite{Vaswani:2017ul} & \begin{tabular}[c]{@{}l@{}}  An encoder-decoder architecture that employs  \\ self-attention  and ANNs in encoder and decoder parts.  \end{tabular} \\ \hline

Neural Machine Translation (NMT) \cite{bahdanau2014neural} &  A seq2seq translation architecture.  \\ \hline

Reinforcement Learning (RL) \cite{sutton2018reinforcement} & \begin{tabular}[c]{@{}l@{}}  A ML algorithm in which an agent  performs  a series of \\  decisions in order to maximize its  rewards.  \end{tabular} \\ \hline

Transfer Learning  \cite{pan2009survey} & \begin{tabular}[c]{@{}l@{}} A methodology to learn a model on a task (or on a large  \\ data) and then to adjust (i.e. fine-tune) the learned model \\ on a  different task (or on a smaller dataset)   with the final \\ goal of generalization.    \end{tabular} \\ \hline

Teacher Forcing  \cite{williams1989learning} & \begin{tabular}[c]{@{}l@{}}  A technique that is used in training RNNs such that the\\ actual word is given to the decoder as the input instead\\ of the  output word that is predicted in the previous step.   \end{tabular} \\ \hline

\end{tabular}}
\end{table}

\begin{table}[H]
\caption{Commonly used databases in drug discovery} \label{tab:resources}
\scalebox{0.72}{
\begin{tabular}{|l|l|l|}
\hline
\textbf{Source}   & \textbf{Address} & \textbf{Description} \\ \hline
UniProt \cite{apweiler2004uniprot}         & https://www.uniprot.org/               & \begin{tabular}[c]{@{}l@{}}The Universal Protein Resource:  stores  protein sequence  and function \\ information. \end{tabular}                                                                               \\ \hline
PDB \cite{berman2000protein} & https://www.rcsb.org/                  & \begin{tabular}[c]{@{}l@{}}The Protein Data Bank: a source of  structural information for around \\ 152,000  macro-molecular structures.\end{tabular}                                                                                                              \\ \hline

PFam \cite{bateman2004pfam}                & https://pfam.xfam.org/                 & \begin{tabular}[c]{@{}l@{}} A protein family database  based on multiple sequence alignment (MSA) \\ and Hidden Markov Models (HMM).\end{tabular}   \\ \hline
PROSITE \cite{hulo2006prosite}             & https://prosite.expasy.org/            & 
\begin{tabular}[c]{@{}l@{}}
A database that contains protein domains, motifs, families and functional \\ sites.  
\end{tabular}
\\ \hline

PubChem \cite{bolton2008pubchem}           & https://PubChem.ncbi.nlm.nih.gov/      & \begin{tabular}[c]{@{}l@{}} An extensive resource for around 96 million compounds and 265 million\\  substances. PubChem also acts as a cheminformatics tool  by providing an \\ interface   that enables the computation of 2D/3D similarity of compounds  \\  and introduces a 1D chemical descriptor.\end{tabular} \\ \hline

ChEMBL \cite{gaulton2011chembl}            & https://www.ebi.ac.uk/chembl/          & \begin{tabular}[c]{@{}l@{}} A widely accessed database that stores manually curated information \\ about  protein targets, chemical properties and bioactivities  for \\ 1.9 million compounds. \end{tabular}    \\ \hline

DrugBank \cite{wishart2006drugbank}        & https://www.drugbank.ca/               & \begin{tabular}[c]{@{}l@{}} An online resource for  chemical, pharmacological and  pharmaceutical \\   information for 13K drugs and 5K proteins (e.g. drug targets/enzymes) \\ that are associated with these drugs. \end{tabular}      \\ \hline

BindingDB  \cite{liu2006bindingdb}         & https://www.bindingdb.org/             &\begin{tabular}[c]{@{}l@{}}  A database of protein and small molecule interactions that   stores \\ binding affinities.      \end{tabular}                                     \\ \hline

PDB-Bind \cite{wang2005pdbbind}            & www.pdbbind.org.cn/                    & A public resource for binding affinity data for protein-ligand complexes.                                       \\ \hline

ZINC \cite{irwin2005zinc}                  & https://zinc.docking.org/              & \begin{tabular}[c]{@{}l@{}} A database of  over 230 million  commercially-available compounds in  \\ 3D form.  \end{tabular}                                        \\ \hline

\multicolumn{3}{l}{All databases were accessed on June 28, 2019.} 
\end{tabular}}
\end{table}

\begin{table}[H]
\caption{Different representations of the drug \textit{ampicillin} } \label{tab:ampicillin}
\begin{tabular}{|l|l|}
\hline
\textbf{Identifier} & \textbf{Representation}  \\ \hline
IUPAC name          & \begin{tabular}[c]{@{}l@{}}(2S,5R,6R)-6-{[}{[}(2R)-2-amino-2-phenylacetyl{]}amino{]}-3,3-\\ dimethyl-7-oxo-4-thia-1-azabicyclo{[}3.2.0{]}heptane-2-carboxylic acid\end{tabular}                                 \\ \hline
Chemical Formula    & $C_{16}H_{19}N_3O_4S$   \\ \hline
Canonical SMILES    & CC1(C(N2C(S1)C(C2=O)NC(=O)C(C3=CC=CC=C3)N)C(=O)O)C  \\ \hline
Isomeric SMILES     & \begin{tabular}[c]{@{}l@{}}CC1({[}C@@H{]}(N2{[}C@H{]}(S1){[}C@@H{]}(C2=O)NC(=O){[}C@@H{]}\\ (C3=CC=CC=C3)N)C(=O)O)C\end{tabular} \\ \hline
\begin{tabular}[c]{@{}l@{}} DeepSMILES \\ (Canonical) \end{tabular}         & CCCNCS5)CC4=O))NC=O)CC=CC=CC=C6))))))N)))))))C=O)O)))C  \\ \hline
\begin{tabular}[c]{@{}l@{}}SELFIES \\ (Canonical)  \end{tabular}        & \begin{tabular}[c]{@{}l@{}}

[C][C][Branch2\_3][Ring1][epsilon][C][Branch2\_3]\\ \textnormal{[epsilon]}[=O][N][C][Branch1\_3][Ring2][S][Ring1][Ring2]\\ \textnormal{[C]}[Branch1\_3][Branch1\_1][C][Ring1][Ring2][=O][N][C]\\ \textnormal{[Branch1\_3]}[epsilon][=O][C][Branch1\_3][Branch2\_2][C][=C]\\ \textnormal{[C]}[=C][C][=C][Ring1][Branch1\_1][N][C][Branch1\_3]\\ \textnormal{[epsilon]}[=O][O][C] \end{tabular}  \\ \hline
InChi               & \begin{tabular}[c]{@{}l@{}}InChI=1S/C16H19N3O4S/c1-16(2)11(15(22)23)19-13\\ (21)10(14(19)24-16)18-12(20)9(17)8-6-4-3-5-7-8/h3-7\\ 9-11,14H,17H2,1-2H3,(H,18,20)(H,22,23)/t9-,10-,11+\\ 14-/m1/s1\end{tabular} \\ \hline
InChi Key           & AVKUERGKIZMTKX-NJBDSQKTSA-N   \\ \hline

2D    & \includegraphics[scale=0.32]{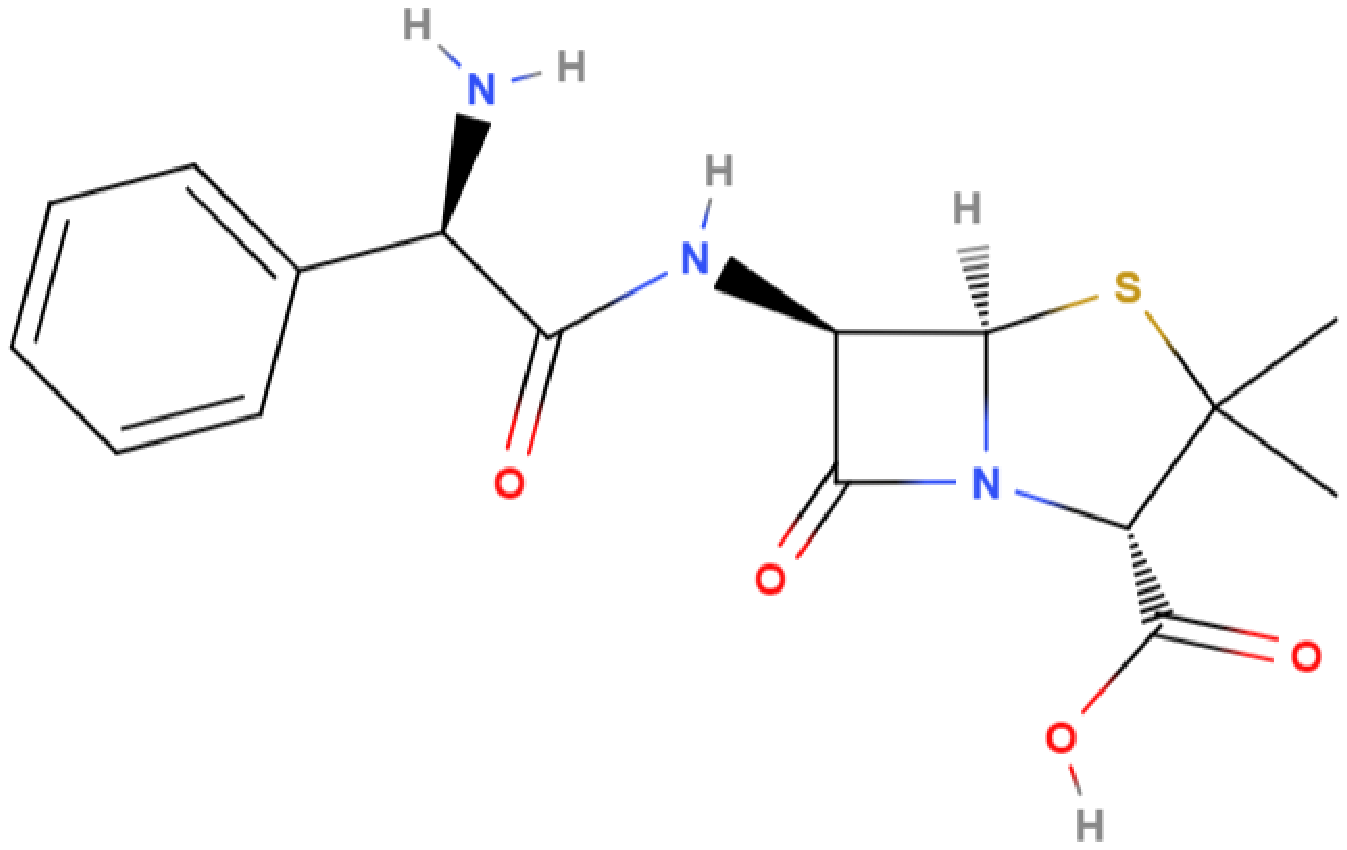}  
\\ \hline

3D    &  \includegraphics[scale=0.32]{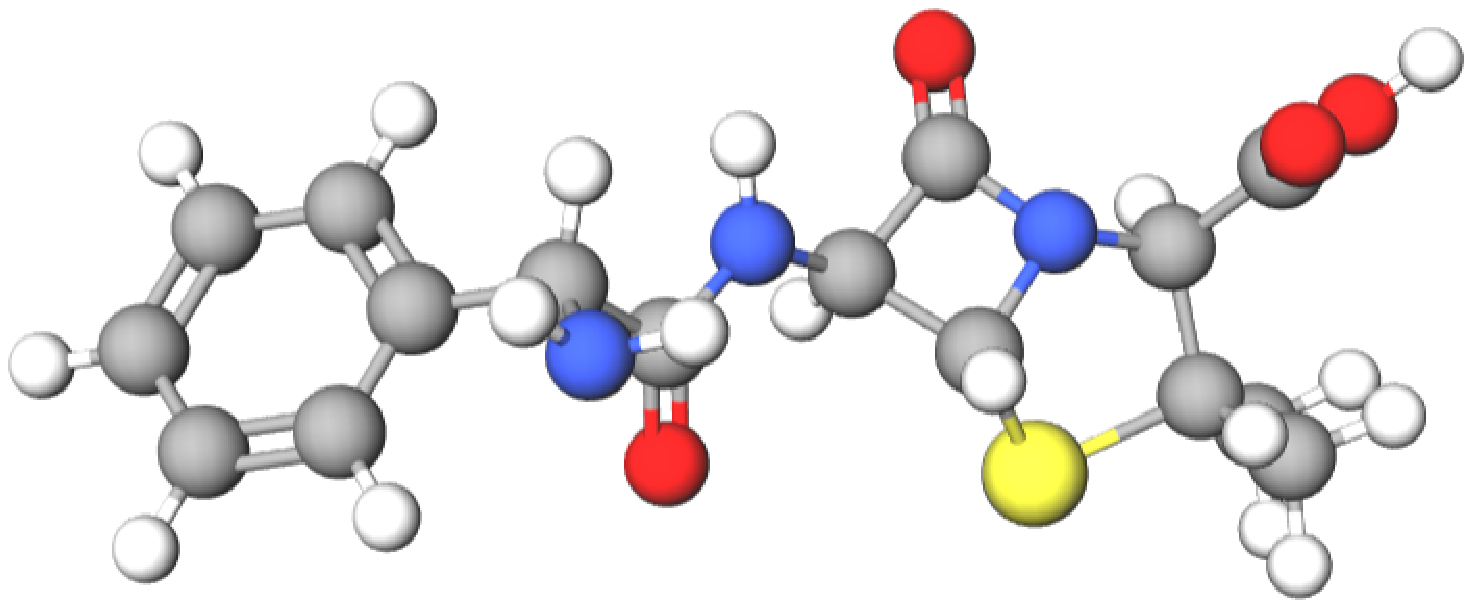}   
\\ \hline

\multicolumn{2}{l}{2D and 3D figures were generated using MolView (\url{molview.org}).} 
\end{tabular}
\end{table}

\end{document}